\documentclass[a4paper,10pt,twoside]{cpc-hepnp}

\usepackage{multicol}
\usepackage{graphicx}
\usepackage{booktabs}
\usepackage{amssymb,bm,mathrsfs,bbm,amscd}
\usepackage[tbtags]{amsmath}
\usepackage{lastpage}

\newcommand*{\no}{\noindent}
\newcommand*{\bea}{\begin{eqnarray}}
\newcommand*{\eea}{\end{eqnarray}}
\newcommand*{\be}{\begin{equation}}
\newcommand*{\ee}{\end{equation}}
\newcommand*{\pd}{\partial}
\newcommand*{\pdm}{\pd_{\mu}}

\newcommand*{\mn}{{\mu\nu}}

\newcommand*{\nn}{\nonumber}

\begin{document}

\fancyhead[co]{\footnotesize A. Maas: Describing gluons at zero and finite temperature}

\footnotetext[0]{Received}

\title{Describing gluons at zero and finite temperature}

\author{Axel Maas\email{Axel.Maas@uni-graz.at}
}
\maketitle

\address{%
Institute of Physics, Karl-Franzens University Graz, Universit\"atsplatz 5, A-8010 Graz, Austria\\
}

\begin{abstract}
Any description of gluons requires a well-defined gauge. This is complicated non-perturbatively by Gribov copies. A possible method-independent gauge definition to resolve this problem is presented and afterwards used to study the properties of gluons at any temperature. It is found that only chromo-electric properties reflect the phase transition. From these the gauge-invariant phase transition temperature is determined for SU(2) and SU(3) Yang-Mills theory independently.
\end{abstract}

\begin{keyword}
Gluons, gauge-fixing, temperature, Yang-Mills theory, Gribov-Singer obstruction, confinement
\end{keyword}

\begin{pacs}
11.15.Ha, 12.38.Aw, 14.70.Dj
\end{pacs}

\begin{multicols}{2}

\section{Gauge-fixing}

The description of any elementary particle, in particular gluons, in the standard model is necessarily gauge-dependent. Their properties are therefore only well-defined after a particular gauge is chosen. In a full non-perturbative calculation gauge-fixing is obstructed due to the presence of Gribov-Singer copies\cite{Gribov}. This ambiguity has to be resolved to obtain results which can be compared across different calculations and methods. However, such a resolution has to necessarily contain non-local components\cite{Gribov}.

To make the definition of such a gauge as accessible by as many methods as possible it would be desirable to define it by imposing conditions on correlation functions. This is possible in perturbation theory, where, e.\ g., Landau gauge can be defined by requiring the longitudinal gluon propagator $p_\mu p_\nu D_\mn$ to vanish.

An investigation of the properties of Gribov copies after imposing the Landau gauge condition $\pdm A_\mu^a=0$ and restricting to the first Gribov horizon by imposing that the operator $M^{ab}=-\pdm(\delta^{ab}\pdm+igf^{abc} A_\mu^c)$ must be positive semi-definite can be performed using lattice gauge theory\cite{Maas:2009se}. On small lattices it has been found that it is sufficient to impose that the ghost propagator $D_G=<M^{-1}>$ must satisfy
\be
B=\frac{p^2D_G(p)}{P^2D_G(P)}\nn
\ee
\no for a chosen fixed value of $B>0$ on the average to resolve the ambiguity. The two momenta $P$ and $p$ are chosen in the perturbative domain and at the lowest accessible momentum, respectively. The range of possible $B$ values depends on the lattice volume and discretization\cite{Maas:2009se}. It remains to be investigated which range is accessible. However, functional studies in the continuum and infinite volume find solutions for any $B$ between a lower positive bound and positive infinity\cite{Boucaud:2008ky,Fischer:2008uz}, giving rise to a decoupling-type and a scaling-type behavior\cite{Fischer:2008uz}, respectively. Since $B$ is a free parameter in the functional calculations, the lattice calculations suggest that it is a gauge parameter, and the different results correspond to different gauges\cite{Maas:2009se,Fischer:2008uz}. However, further investigations are necessary to establish whether this interpretation is correct. This will be assumed here henceforth.

\section{Finite temperature}

At finite temperature the gluon propagator must be described in terms of two independent dressing functions
\be
D_\mn=P_\mn^T D_T(p)+P_\mn^L D_L(p)\nn,
\ee
\no with the projectors $P_\mn^T$ transverse w.\ r.\ t.\ the heat-bath and $P_\mn^L$ longitudinal. Selecting a gauge in which a scaling-type behavior is enforced, it is possible to investigate the properties of the two dressing functions $D_T$ and $D_L$ with functional methods\cite{Cucchieri:2007ta,Maas:2004se}. It is found that the transverse function $D_T$ vanishes for any gauge group at all temperatures\cite{Cucchieri:2007ta}. As a consequence, it cannot be described by a positive spectral function, and transverse gluons are confined at all temperatures\cite{Maas:2004se}.

\end{multicols}
\begin{center}
\includegraphics[width=0.5\linewidth]{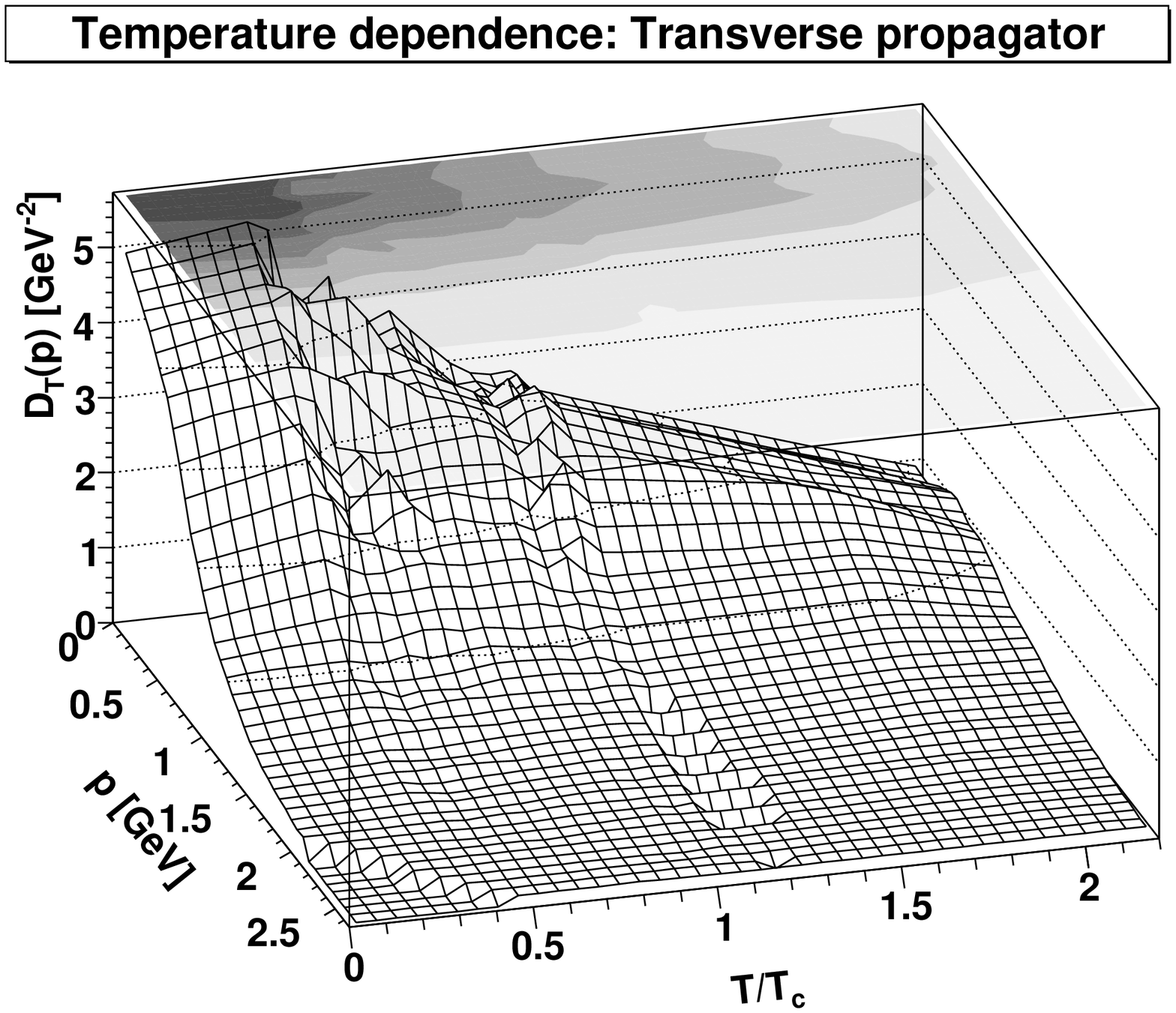}0.5\includegraphics[width=0.5\linewidth]{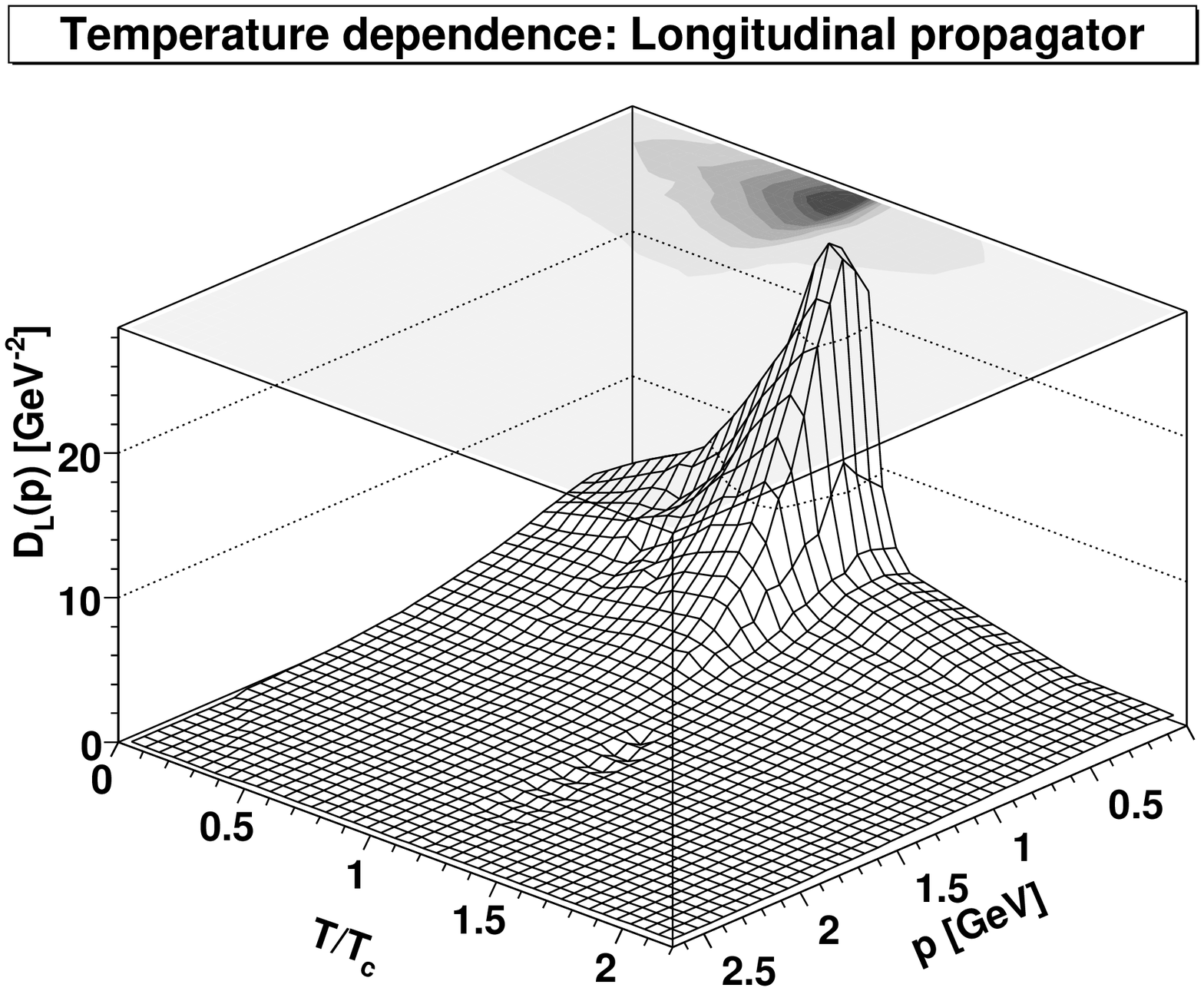}
\includegraphics[width=0.5\linewidth]{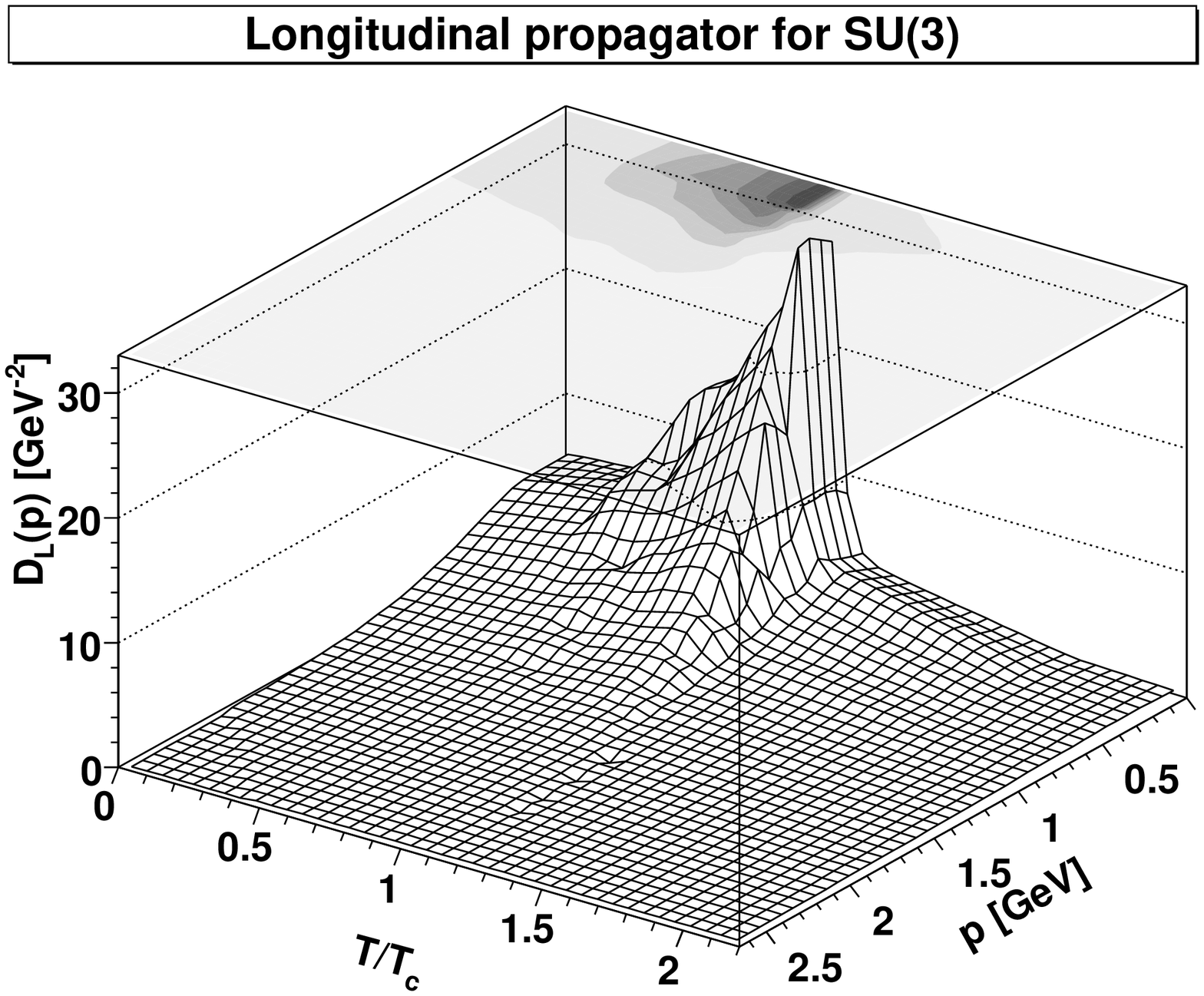}0.5\includegraphics[width=0.5\linewidth]{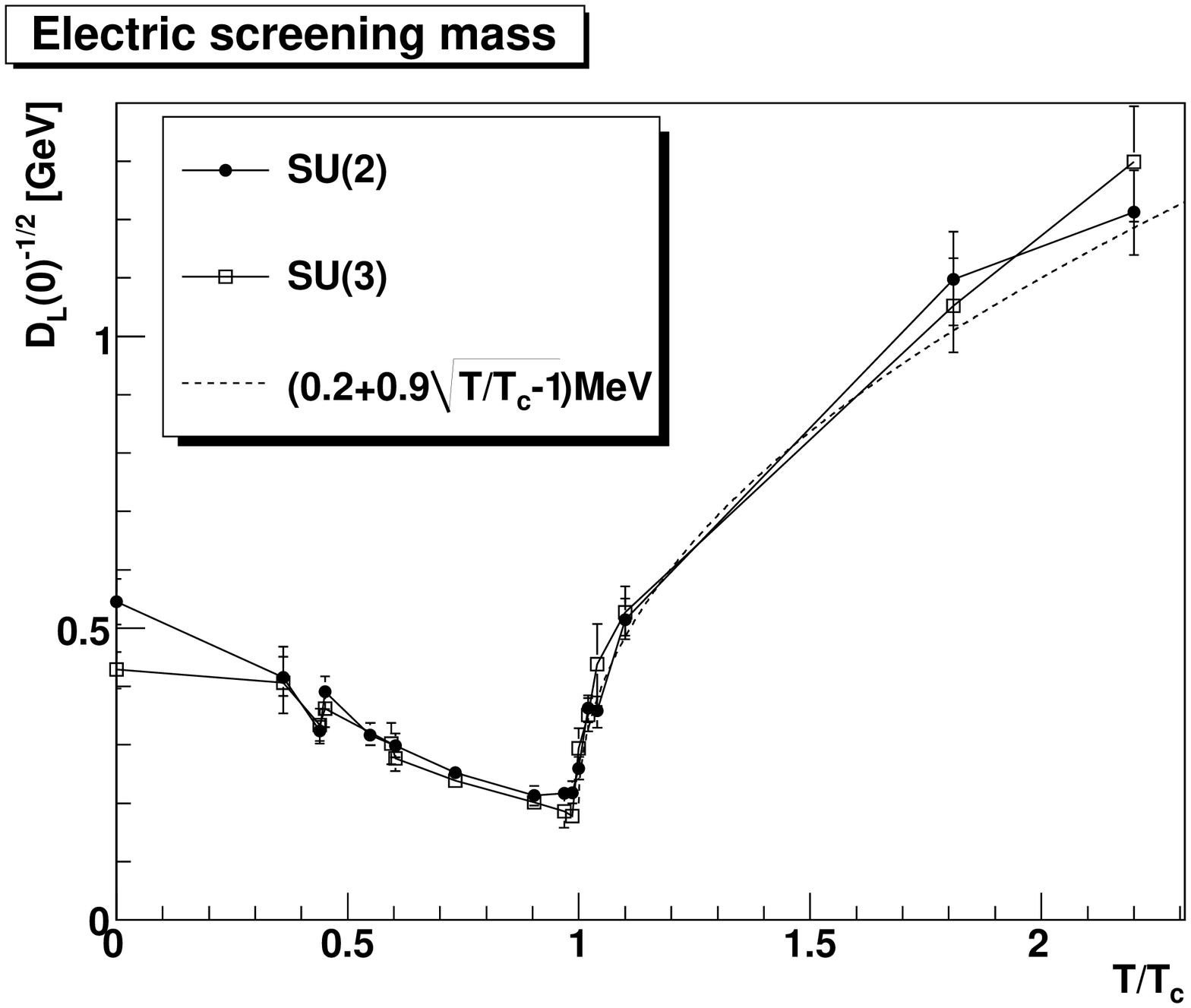}
\figcaption{\label{res} Top left panel: Transverse gluon propagator $D_T$ for SU(2) as a function of temperature and momentum. Top right panel: Longitudinal gluon propagator $D_L$ for SU(2) as a function of temperature and momentum. Bottom left panel: Longitudinal gluon propagator $D_L$ for SU(3) as a function of temperature and momentum. Bottom right panel: Electric screening mass for SU(2) and SU(3) together with a fit of the high-temperature domain. Volumes are between (3.5 fm)$^4$ at zero temperature and (9.4 fm)$^4$ at the highest temperature, with $a$ ranging between 0.2 and 0.16 fm. Details will be available elsewhere{\protect\cite{more}}.}
\end{center}
\begin{multicols}{2}

This also applies to gluons polarized longitudinally w.\ r.\ t.\ the heat-bath, as they belong already perturbatively to a BRST quartet\cite{Cucchieri:2007ta}. Their propagator is dominated at low momenta by an electric screening mass\cite{Cucchieri:2007ta,Cucchieri}. It emerges because the longitudinally polarized gluon ceases almost completely to interact ultra-softly, in contrast to the transversely polarized one\cite{Maas:2005hs}. Therefore, it is only influenced by interactions with hard modes, which provide a screening mass on the order of the temperature.

Both facts together imply that gluons are confined at all temperatures. But this is not in contradiction to a Stefan-Boltzmann-like behavior of thermodynamic quantities, as the latter are dominated by hard interactions, and the confining interactions are thermodynamically sub-leading at large temperatures\cite{Maas:2004se,Maas:2005hs}.

However, it is not yet possible to determine the temperature-behavior of the screening mass using functional methods\cite{Cucchieri:2007ta}, but see\cite{Fischer:2008uz}. For this purpose here lattice gauge theory is used\footnote{Here actually a decoupling-type gauge is employed, but the difference at presently accessible volumes and discretizations for the gluon propagator are yet negligible\cite{Maas:2009se}.}. The results, using the methods described in\cite{Cucchieri:2007ta,Maas:2007af}, are shown in figure \ref{res}, top panels, for SU(2) Yang-Mills theory.

The left panel shows the transverse propagator as a function of temperature. It is weakly temperature-dependent, in particular not reflecting the phase transition. This is expected, since the ultra-soft interactions dominating its low momentum regime are stronger than the ones on the order of the temperature\cite{Maas:2005hs}. In particular, as no deconfinement occurs, the phase transition is not leaving an imprint. This is different for the hard-mode-dominated longitudinal propagator\cite{Cucchieri:2007ta}, as shown in the top-right panel of figure \ref{res}. It is visible that it strongly reacts to the phase transition. Analyzing the corresponding electric screening mass in the bottom-right panel of figure \ref{res} shows that it sensitively reacts to the phase transition, decreasing below and quickly increasing, proportional to $\sqrt{T}$, above. The transition is sharp enough for the independent determination of the phase transition temperature.

Naturally the question arises whether there is a difference for the second and first order phase transitions from SU(2) and SU(3). In fact, the situation is similar in SU(3): The transverse sector shows so little difference to SU(2) that it is not shown\cite{more}. In the bottom-left panel of figure \ref{res} the result for the longitudinal sector is displayed, showing a very similar behavior as for SU(2). The electric screening mass, also shown in the bottom-right panel of figure \ref{res}, is very similar, though the SU(3) results appear minutely more spiky at the transition. If the order of the phase transition is therefore leaving an imprint in these correlation functions will hence be due to whether a scaling or jumping behavior of the electric screening mass at the phase transition is observed in the thermodynamic limit. To decide this requires a careful and detailed analysis in the future.

\section{Summary}

The description of gluons is necessarily gauge-dependent. To be able to compare the results of different methods an unambiguous definition of the gauge is required, which is non-perturbatively difficult due to Gribov copies. A proposal for such a non-perturbative gauge-fixing is based on imposing conditions on the Landau-gauge ghost propagator\cite{Maas:2009se}. Though this proposal requires very much further investigations, it could unite, by means of a second gauge parameter, all presently available results at zero\cite{Boucaud:2008ky,Fischer:2008uz,other} and finite temperature\cite{Cucchieri,Cucchieri:2007ta,Saito:2009bn} on Landau-gauge propagators.

Assuming this to be correct, and selecting a suitable gauge, it has been found that gluons are not deconfined at all temperatures\cite{Cucchieri:2007ta}, without contradicting the Stefan-Boltzmann behavior of thermodynamic quantities\cite{Maas:2005hs}. The results presented here show furthermore that the phase transition leaves its imprint in the electric screening mass, for both first and second order phase transition, i.\ e., for SU(2) and SU(3) Yang-Mills theory, respectively. In fact, the imprint is sufficiently strong for an independent determination of the gauge-invariant transition temperature using the gauge-dependent gluon propagators. Whether the temperature dependence of the electric screening mass is also containing information on the order of the phase transition has to be investigated carefully in the future.

\acknowledgments{This work was supported by the FWF under grant number M1099-N16.}

\end{multicols}

\vspace{-2mm}
\centerline{\rule{80mm}{0.1pt}}
\vspace{2mm}

\begin{multicols}{2}

\end{multicols}

\vspace{5mm}

\clearpage


\begin{thebibliography}{90}

\vspace{3mm}

\bibitem{Gribov}
  V.\ N.\ Gribov,
  Nucl.\ Phys.\ B {\bf 139}, 1 (1978);
  I.\ M.\ Singer,
  Commun.\ Math.\ Phys.\ {\bf 60} (1978) 7.

\bibitem{Maas:2009se}
  A.~Maas,
  arXiv:0907.5185 [hep-lat];
  in progress.

\bibitem{Boucaud:2008ky}
  Ph.~Boucaud, et al.\
  JHEP {\bf 0806}, 099 (2008)
  [arXiv:0803.2161].

\bibitem{Fischer:2008uz}
  C.~S.~Fischer, A.~Maas and J.~M.~Pawlowski,
  Annals Phys.\  {\bf 324} (2009) 2408
  [arXiv:0810.1987 [hep-ph]].

\bibitem{Cucchieri:2007ta}
  A.~Cucchieri, A.~Maas and T.~Mendes,
  Phys.\ Rev.\  D {\bf 75} (2007) 076003
  [arXiv:hep-lat/0702022].
  
\bibitem{Maas:2004se}
  A.~Maas, J.~Wambach, B.~Gr\"uter and R.~Alkofer,
  Eur.\ Phys.\ J.\  C {\bf 37} (2004) 335
  [arXiv:hep-ph/0408074].

\bibitem{Cucchieri}
A.\ Cucchieri, F.\ Karsch and P.\ Petreczky,
Phys.\ Lett.\ B {\bf 497}, 80 (2001)
[arXiv:hep-lat/0004027];
Phys.\ Rev.\ D {\bf 64}, 036001 (2001)
[arXiv:hep-lat/0103009].

\bibitem{Maas:2005hs}
  A.~Maas, J.~Wambach and R.~Alkofer,
  Eur.\ Phys.\ J.\  C {\bf 42} (2005) 93
  [arXiv:hep-ph/0504019].

\bibitem{Maas:2007af}
  A.~Maas and {\v S}.~Olejn\'ik,
  JHEP {\bf 0802} (2008) 070
  [arXiv:0711.1451 [hep-lat]].

\bibitem{more}
A.~Maas, work in progress.

\bibitem{other}
  A.~Cucchieri and T.~Mendes,
  Phys.\ Rev.\  D {\bf 78} (2008) 094503
  [arXiv:0804.2371 [hep-lat]];
  Phys.\ Rev.\ Lett.\  {\bf 100} (2008) 241601
  [arXiv:0712.3517 [hep-lat]];
  V.~G.~Bornyakov, V.~K.~Mitrjushkin and M.~M\"uller-Preussker,
  Phys.\ Rev.\  D {\bf 79} (2009) 074504
  [arXiv:0812.2761 [hep-lat]];
  D.~Dudal, J.~A.~Gracey, S.~P.~Sorella, N.~Vandersickel and H.~Verschelde,
  Phys.\ Rev.\  D {\bf 78} (2008) 065047
  [arXiv:0806.4348 [hep-th]];
  A.~Sternbeck and L.~von Smekal,
  arXiv:0811.4300 [hep-lat];
  I.~L.~Bogolubsky, E.~M.~Ilgenfritz, M.~M\"uller-Preussker and A.~Sternbeck,
  Phys.\ Lett.\  B {\bf 676} (2009) 69
  [arXiv:0901.0736 [hep-lat]];
  O.~Oliveira and P.~J.~Silva,
  arXiv:0910.2897 [hep-lat];
  D.~Binosi and J.~Papavassiliou,
  Phys.\ Rept.\  {\bf 479} (2009) 1
  [arXiv:0909.2536 [hep-ph]].

\bibitem{Saito:2009bn}
  T.~Saito, M.~N.~Chernodub, A.~Nakamura and V.~I.~Zakharov,
  arXiv:0910.4828 [hep-lat].

\end{thebibliography}
\end{document}